\documentclass[aps,epsfig,nofootinbib]{revtex4} 
\usepackage{graphicx}
\input epsf
\newcommand{\sfig}[2]{
\centerline{ \includegraphics[width=#2]{#1} }
		}
\newcommand{\Sfig}[2]{
	\begin{figure}[thp]
	\sfig{#1.eps}{0.85\columnwidth}
	\caption{{\small #2}}
	\label{fig:#1}
	\end{figure}
}
\newcommand{\Sfigr}[2]{
	\begin{figure}[thp]
\centerline{\includegraphics[width=0.85\columnwidth,angle=-90]{#1.eps} }
	\caption{{\small #2}}
	\label{fig:#1}
	\end{figure}
}

\newcommand{\Rf}[1]{\ref{fig:#1}}

\def\cmm2{{\,\rm cm^{-2}}}
\def\cm2{{\,{\rm cm}^2}}
\def\cmm3{{\,{\rm cm}^{-3}}}
\def\gcmm3{{\,{\rm g\,cm^{-3}}}}

\def\la{\mathrel{\mathpalette\fun <}}

\def\fun#1#2{\lower3.6pt\vbox{\baselineskip0pt\lineskip.9pt
  \ialign{$\mathsurround=0pt#1\hfil##\hfil$\crcr#2\crcr\sim\crcr}}}

\def\be{\begin{equation}}
\def\ee{\end{equation}}
\def\bea{\begin{eqnarray}}
\def\eea{\end{eqnarray}}
\newcommand{\vs}{\nonumber\\}

\newcommand{\ec}[1]{equation~(\ref{eq:#1})}

\newcommand{\eql}[1]{\label{eq:#1}}
\newcommand{\spv}{\mathcal{V}}

\begin{document}

\title{Everpresent $\Lambda$} 

\author{Maqbool Ahmed$^{1}$, Scott Dodelson$^{2,3}$, Patrick B. Greene$^{2}$,
and Rafael Sorkin$^{1}$
}

\affiliation{$^1$Department of Physics, Syracuse University,Syracuse, NY
13244-1130}
\affiliation{$^2$NASA/Fermilab Astrophysics Center,
Fermi National Accelerator Laboratory, Batavia, IL~~60510-0500}
\affiliation{$^3$Department of Astronomy \& Astrophysics, The University of
Chicago, Chicago, IL~~60637-1433}

\date{\today}

\begin{abstract}

A variety of observations indicate that the universe is dominated by
dark energy with negative pressure, one possibility for which is a
cosmological constant.  If the dark energy is a cosmological constant, a
fundamental question is: Why has it become relevant at so late an epoch,
making today the only time in the history of the universe at which the
cosmological constant is of order the ambient density.  We explore an
answer to this question drawing on ideas from unimodular gravity, which
predicts fluctuations in the cosmological constant, and causal set
theory, which predicts the magnitude of these fluctuations.  The
resulting ansatz yields a fluctuating cosmological ``constant'' which is
always of order the ambient density.

\end{abstract}
\maketitle

\section{Introduction}

The most startling discovery to emerge from the recent plethora of
cosmological data is that the universe appears to be dominated by dark
energy~\cite{Concordance,SN,CMB}.  We know that this dark energy accounts
for roughly seventy percent of the energy density in the universe, does not
cluster like ordinary matter, and has negative pressure. Otherwise, we
are in the dark about the nature of this extraordinary phenomenon. 

Perhaps the most popular explanation is that the dark energy is due to
a cosmological constant, for such a parameter was introduced into general
relativity at its birth~\cite{Einstein} and has remained an important
tool for cosmologists seeking to model the observed universe~\cite{age}. 
The strongest argument against the cosmological constant is that naively
we expect it to contribute an energy density, $\rho_\Lambda$, of order
$m_p^4 = (8\pi G)^{-2}$, where $G$ is Newton's constant and $m_p$ is
the reduced Planck Mass\footnote%
{In this paper, we will use units in which $\hbar=c=m_p=1$.}.
This estimate is some one hundred and twenty orders of magnitude
larger than the observed value.  An equivalent way of stating the problem
is to note that only today is the cosmological constant of order the ambient
density in matter or radiation.   At all past epochs, $\rho_\Lambda$ was
sub-dominant and immensely so.
Many people have felt that no theory could naturally predict
such a tiny value for $\Lambda$ (or equivalently such a late epoch for
it to become relevant) without predicting $\Lambda$ to vanish entirely,
and for this reason they have sought other explanations of the observations.

Many alternatives to the cosmological constant have been proposed.  Most
significant among these are quintessence models in which the dark energy
is due to a homogeneous scalar field shifted away from the true
minimum of its potential \cite{quintessence}.   
Like a simple cosmological constant, many of these suffer from the ``Why
Now?'' problem: Why does the quintessence field come to dominate only
recently?
They also typically need to explain the
small mass scale necessary for the field to be important today
($m \la 10^{-33}$ eV).   Even more disturbing, none of them are connected
to realistic particle physics models.  Perhaps then, instead of altering
the energy content of the universe, we need to look in another direction
and modify gravity in order to explain the dark energy today.

The simulations reported here flesh out an old heuristic
prediction~\cite{LamPred} of a fluctuating cosmological term arising from the
basic tenets of causal set theory.\footnote%
{For an introduction to the causal set hypothesis see~\cite{prlcs}.}
In normal usage, the words ``cosmological term''
refer to a contribution to the effective
stress-energy-momentum tensor of the form
$T_{\mu \nu} = g_{\mu \nu} \Lambda(x)$.  
However, in classical General Relativity (GR) such a $\Lambda(x)$
must be constant if the total energy momentum in other
matter components is separately conserved.  Here we consider a
specific modification of GR motivated by the search for a theory 
of quantum gravity based on causal sets.

Although the ultimate status and precise interpretation of the 
prediction of a fluctuating $\Lambda$ 
must await the development of a
quantum dynamics for causal sets~\cite{dprrds}, 
the basic lines of the
argument are simple and general enough that they have a certain
independence of their own.  
In this paper we review the motivation for a
fluctuating $\Lambda$ from causal set theory, propose an ansatz for the
form of these fluctuations, apply the latter to the Friedmann equation
with time-dependent cosmological term, and find that we can have a
viable cosmology for some fraction of the solutions.  Finally, we
address issues related to our choice of evolution equations.

\section{Causet Theory}

Here, we will review the arguments leading to a fluctuating cosmological
term and then describe the specific ansatz via which we have chosen to
implement their main implication: that $\Lambda$ can be expected to
fluctuate and with a magnitude that diminishes as the universe grows
older.

In causal set (``causet'') theory, the predicted fluctuations arise, as
a kind of residual (and nonlocal) quantum effect, from the underlying
space-time discreteness.
More specifically, the basic inputs to the argument are: 
space-time discreteness leading to a finite number $N$ of elements; 
the interpretation of space-time volume $\spv$ as a direct reflection of $N$; 
the conjugacy of $\Lambda$ to space-time volume $\spv$;
and the 
existence of fluctuations in $\spv$ coming from Poisson fluctuations in $N$.  
(Of these four inputs, the first is not peculiar to causal sets, but the
remaining ones all are to a greater or lesser extent.)

The two most basic tenets of causal set theory are first, that the causal
ordering of events in macroscopic space-time reflects a more fundamental
order relation among the elements of an underlying discrete structure to
which continuous spacetime is only an approximation, and second, that the
four-volume of a region of spacetime reflects the number of discrete
elements of which the region is ``composed''.
The hypothesized discrete substratum or {\it causal set} is taken to be a
partially ordered set and its dynamics is conceived of as a kind of
growth process in which elements come into being, one at a time. 
Although a classically stochastic dynamics expressing these ideas is by
now fairly well developed~\cite{developed}, a corresponding quantum dynamics is
only just beginning to be sought.  Any prediction of quantum fluctuations in
$\Lambda$ must therefore rest on an anticipation of certain features of
this ``new QCD'' (quantum causet dynamics).   

Let us begin by assuming that, 
at some level of approximation, this
dynamics will correspond to a space-time ``path integral'' in which one
is summing over certain classes of four-geometries.  At the deeper level
however, this will still be a sum over causal sets.  
Then let us take
from the already developed classical growth models for causal sets the
feature that the ever growing number $N$ of causet elements plays the
role of a kind of parameter time -- the time in which the stochastic
process which mathematically represents the growth unfolds, and with
respect to which the probabilities are normalized.  Just as one does not
sum over time in ordinary quantum mechanics, one would not expect to sum
over causets with different values of $N$ in the quantum theory.  But,
because number corresponds macroscopically to volume $\spv$, this
translates into the statement that one should hold $\spv$ fixed in
performing the gravitational path integral.  Any wave function that
arises will therefore depend not only on suitable boundary data (say a
three-geometry) but also on a four-volume parameter $\spv$.  Such a
restricted path integral may be called ``unimodular''.

Now the unimodular modification of ordinary GR has been
fairly well studied~\cite{unimodular}, and it is understood that within it,
$\Lambda$ and $\spv$ are conjugate in the same way that energy and time are
conjugate in ordinary quantum mechanics.  (Indeed this is almost obvious
from the fact that the cosmological constant term in the action-integral of
general relativity is just the product $-\Lambda \spv$.)  In particular,
this means that, to the extent that $\spv$ is held fixed in the
gravitational sum-over-histories, $\Lambda$ will be entirely
undetermined by the fundamental parameters of the theory.  (Again this
is almost obvious by reference to the classical limit of unimodular
gravity, where the Lagrange multiplier used to implement the fixed $\spv$
constraint combines with any ``bare'' $\Lambda$ in such a way that the
observed or ``renormalized'' $\Lambda$ represents nothing more than a
constant of integration.)

If this were the whole story, then our conclusion would be that
$\Lambda$ is subject to quantum fluctuations (just like energy $E$ in
ordinary quantum mechanics) but it would not be possible to say anything
about their magnitude,  nor about the magnitude of the mean $\Lambda$
about which the fluctuations would occur.

But here there enters a second aspect of the causal set hypothesis that
we have not mentioned earlier.  In order to do justice to local Lorentz
invariance, the correspondence between number and volume cannot be exact,
but it must be subject to Poisson type fluctuations\footnote%
{More specifically, the correspondence between the underlying causet and
 the approximating space-time is via a notion of ``Poisson sprinkling''
 at unit density, see references~\cite{prlcs,lucadavid} for details.},
which of course
have a typical scale of $\sqrt{N}$.	
This means that, in holding $N$ fixed at the fundamental level, we in
effect fix $\spv$ only up to fluctuations of magnitude $\pm\sqrt{\spv}$.
(Notice that these are not dynamical fluctuations.  Rather they occur at
a {\it kinematic} level: that of the correspondence between order
theoretic and spatio-temporal variables.)  Hence, we do end up
integrating over some limited range of $\spv$ after all, and
correspondingly we do determine $\Lambda$ to some degree --- but only
modulo fluctuations that get smaller as $\spv$ gets larger.  Specifically,
we have
\begin{equation} \label{dL}
    \Delta\Lambda \sim 1 / \Delta{\spv} \sim 1/ \sqrt{\spv} \, .
\end{equation}

As any proper dilemma should, that of the cosmological constant has two
horns: Why is $\Lambda$ so nearly zero and Why is it not exactly zero?
None of what we have said so far bears on the first question, only on
the second.  All we can conclude is that partially integrating over
$\spv$ in the effective gravitational path integral will drive us toward
{\it some} value of $\Lambda$.  We must assume, as the evidence
overwhelmingly suggests, that this ``target value'' is zero, for reasons
still to be understood.\footnote%
{One possible mechanism is that only $\Lambda=0$ is stable against the
 destructive interference induced by {\it non-manifold} fluctuations of
 the causal set.}
Then we end up predicting fluctuations about zero of a magnitude given
by~(\ref{dL}).

Independent of specifics, the space-time volume $\spv$
should be roughly equal to  the fourth power of the Hubble radius,
$H^{-1}$.  Therefore, at all times we expect the energy density in the
cosmological constant to be of order
\be
    \rho_\Lambda \sim \spv^{-1/2} \sim H^2 \sim \rho_{\rm critical}
\ee
the critical density (recall that we are setting $8\pi G=1$).
We thus 
obtain a prediction for today's $\Lambda$
which agrees in order of magnitude with current fits to the astronomical
data.  
And this argument is not limited to today: at all times we 
expect the energy density in the cosmological constant to be of order
the critical density.

This is the basic idea, but any attempt to implement it immediately
raises questions whose answers we can at present only guess at, pending
the development of a fuller quantum dynamics for causets.  At a
conceptual level there is first of all the question of precisely how to
interpret the $\spv$ that figures in equation~(\ref{dL}) and second of
all the question how to incorporate a fluctuating $\Lambda$ into some
suitable modification of the Einstein equations.  At a more practical
level, if we aim to understand, for example, how fluctuations in
$\Lambda$ would have affected structure formation, we need to know, not
only their typical magnitude at each moment of cosmic time, but also how
the fluctuations at one moment correlate with those at other moments.

Concerning the conceptual questions, we will, for present purposes,
resolve them provisionally as follows.
First we impose spatial homogeneity, so that the Einstein equations
reduce to a pair of ordinary differential equations for the scale factor
$a$.  
Of 
these two 
equations, 
one, the so called {\it Friedmann equation} or {\it Hamiltonian constraint}, 
is first order in time and embodies the energy law in this setting, while 
the second involves $\ddot{a}$ and, 
in the case of a {\it non-fluctuating} $\Lambda$, 
adds no information to the first, 
except at moments when $\dot{a}=0$.
They cannot both be compatible with a time dependent cosmological term
when other energy momentum components are separately conserved, 
so we choose one over the other.  
Specifically, we choose to interpret $\Lambda$ via 
its role in the Friedmann equation.  
That is, we retain the Friedmann equation but let
$\Lambda$ be a function of time, dropping the second equation
entirely.

\Sfigr{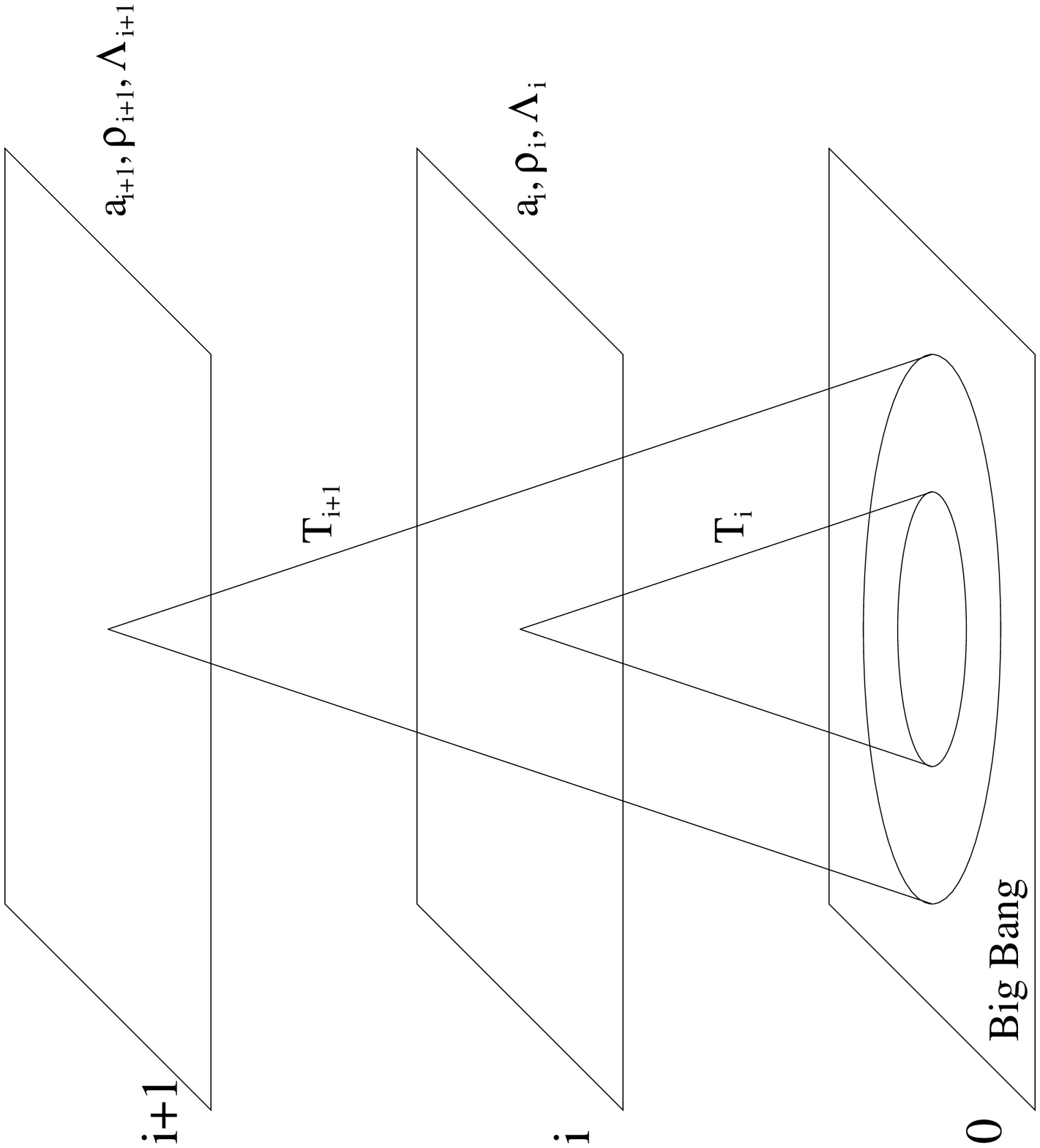}{Schematic representation of the backward
light-cone at two different cosmic times.  Evolution of the scale
factor between the two time slices is determined by the Friedmann equation
while $\Lambda$ varies stochastically.}

The quantity $\spv$ which governs the magnitude of the fluctuations in
$\Lambda$ we will identify (up to an unknown factor of order unity) with
the volume of the past light cone of any representative point on the
hypersurface for which we want the value of $\spv$, as illustrated in
Figure~\Rf{forward9}.  
Although this interpretation is somewhat at odds
with the meaning that $\spv$ has in the unimodular context, it seems
more in accord with causality,
and it is the only number accessible to observation in any useful sense.

With these choices made, the only remaining question is what sort of
random process we want to use to simulate our fluctuating $\Lambda$.
Ideally perhaps, this would be some sort of ``quantal stochastic
process'' (since the underlying process is quantal), but here we do the
simplest thing possible and let the fluctuations in $\Lambda$ be driven
by those of an unadorned random walk.  In fact the ansatz we will use
has some appeal in its own right as an independent ``story'' of why the
cosmological constant might be expected to fluctuate in any discrete
quantum gravity theory that incorporates the equality $N=\spv$ between
volume and number of elements.

With reference to the Einstein-Hilbert Lagrangian, one could describe
the cosmological constant as the ``action per unit space-time volume
which is due just to the existence of space-time as such, independent
of any excitations such as matter or gravitational waves''.
Re-interpreting volume as number of elements, we can say then that
$\Lambda$ is the ``action per element''.  One would expect this to be of
order unity in fundamental units, and if we identify the latter with
Planck units, we get the old answer which is off by some 120 orders
of magnitude.  On the other hand, if we suppose that each element makes
its own contribution and these contributions fluctuate in sign\footnote{It
would probably be more suitable to speak not in terms of action $S$
but rather $\exp(iS/\hbar)$ and say that the contributions (now multiplicative
rather than additive) fluctuate in {\it phase}.}
then the relative smallness of $\Lambda$ will be explained; but one
would also expect a residual $\sqrt{N}$ contribution to $S$ to remain
uncanceled.   Consequently, there would remain a residual contribution
to the action per element of ${\sqrt{N}}/{N} = {1}/{\sqrt{N}}$,
in agreement with our earlier argument.

To implement such an ansatz is now straightforward.  What we need for
the sake of the Friedmann equation is just $\Lambda$ as a function of
$N$ (or equivalently of $\spv$).  To produce such a function we just
generate a string of random numbers of mean 0 and standard deviation 1
(say) and identify $\Lambda(N)$ with the ratio $S(N)/N$, where $S(N)$ is
the sum of the first $N$ of our random numbers.  Modulo implementational
details this is the scheme we have used in the simulations on which we 
report next.

\section{Simulations}

We take as the space-time volume
\be
\spv(t) = {4\pi\over 3} \int_0^t dt' a(t')^3
		\left[ \int_{t'}^t dt''/a(t'') \right]^3
\ee
where $a(t)$ is the scale factor of the universe at proper time $t$. 
Note from this formula that the backward light-cones depicted in
Figure~\Rf{forward9} are quite deceptive: 
because $a(t)$ was much smaller in
the past and vanishes at the Big Bang, 
most of the four-volume $\spv$ of these light cones accumulates
recently.  
One consequence of this is that
$\spv \sim H^{-4}$ recently even if there was a period of cosmic inflation
in the early universe.

Our algorithm for calculating the cosmological constant at time-step $i+1$
is then to set 
\be
\delta N_i \equiv N_{i+1} - N_i = \spv(t_{i+1}) - \spv(t_i)
\ee
and then write
\bea
 \rho_{\Lambda,i+1} &=& { S_{i+1} \over N_{i+1} }
 \vs
  &=& {S_i + \alpha \xi_{i+1} \sqrt{\delta N_i} \over N_i + \delta N_i}
 .\eql{defalp}
\eea
Here 
$\alpha$ is an unknown dimensionless parameter 
which governs the dynamics of the theory; 
$\xi_{i+1}$ is a random number 
with mean 0 and standard deviation 1; 
and $S_0$ is set to zero at some very early time $t_0$. 
We then expand the universe according to
\be
H^2 = \left( {{\dot a} \over a} \right)^2
= {1\over 3} \left( \rho_{\rm matter} + \rho_{\rm radiation} + \rho_\Lambda
\right),
\eql{hubble}\ee
recompute the new space-time volume and repeat.

Figure~\Rf{rhost} shows the evolution of
the energy density in one such realization. 
During the radiation era, $\rho_\Lambda$ scales roughly as $a^{-4}$, 
while during the matter era it scales as $a^{-3}$. 
Thus at all times it is comparable to the ambient energy density. 
If the recipe we have devised for 
implementing
the ideas of causal set theory
and unimodular gravity is an accurate approximation to the
ultimate quantum theory, then these modifications of GR do indeed
lead to an {\it Everpresent $\Lambda$}, a cosmological term which
is always with us~\cite{tracker}.

\Sfig{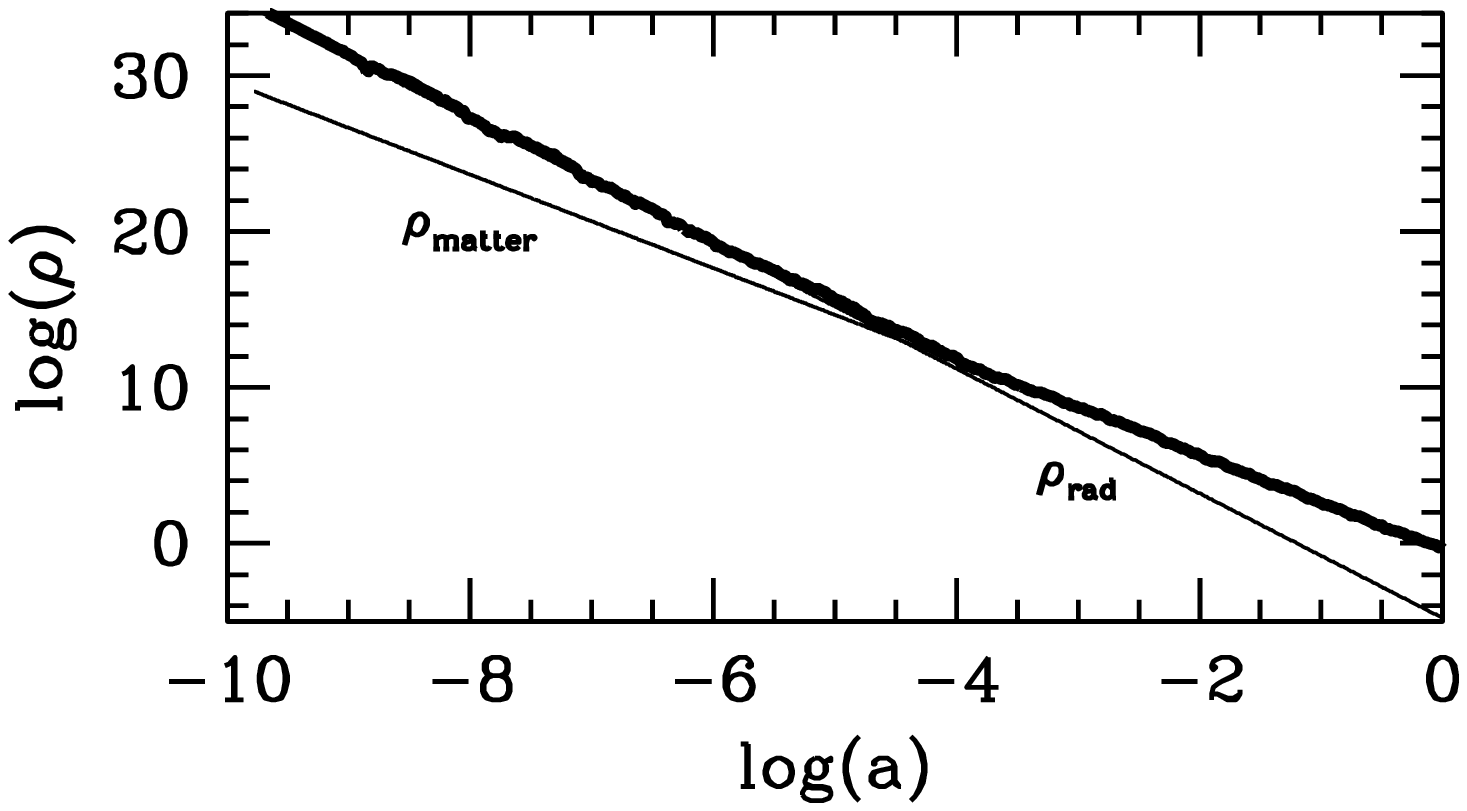}{Evolution of the energy densities in the universe.  The
thick curve is the absolute value of the
energy density in the cosmology constant. The
fluctuating $\rho_\Lambda$ is always of order the ambient density, be it
radiation (early on) or matter (later). Here the dimensionless parameter
$\alpha$ which governs the amplitude of the fluctuations has been set to
$0.01$.}

Hidden in the gross structure of Figure~\Rf{rhost} are the fluctuations
about this average scaling. These fluctuations are crucial if the theory
is to describe the real universe for two reasons: First, there cannot
be too much excess energy at $a\sim 10^{-9}$ or else the successful predictions
of Big Bang Nucleosynthesis (BBN) will be destroyed. 
Second, if $\rho_\Lambda$
scales exactly
as matter today, it will not have the correct equation of
state to account for the cosmological observations. Figure~\Rf{zvt} shows the
ratio of the energy density in $\Lambda$ to the total energy density
as a function of the scale factor for another realization, this time with a
slightly larger value of $\alpha$. This ratio, $\Omega_\Lambda$, fluctuates
about zero with an amplitude of order $0.5$ (as we will shortly see, this
amplitude is a function of $\alpha$). In this particular realization,
$\Lambda$ accounts for over fifty percent of the energy density today
and changes very little going back to redshift $z=1$ ($a=0.5$); 
thus it behaves recently as a true cosmological constant, and therefore
satisfies the observed cosmological constraints.

\Sfig{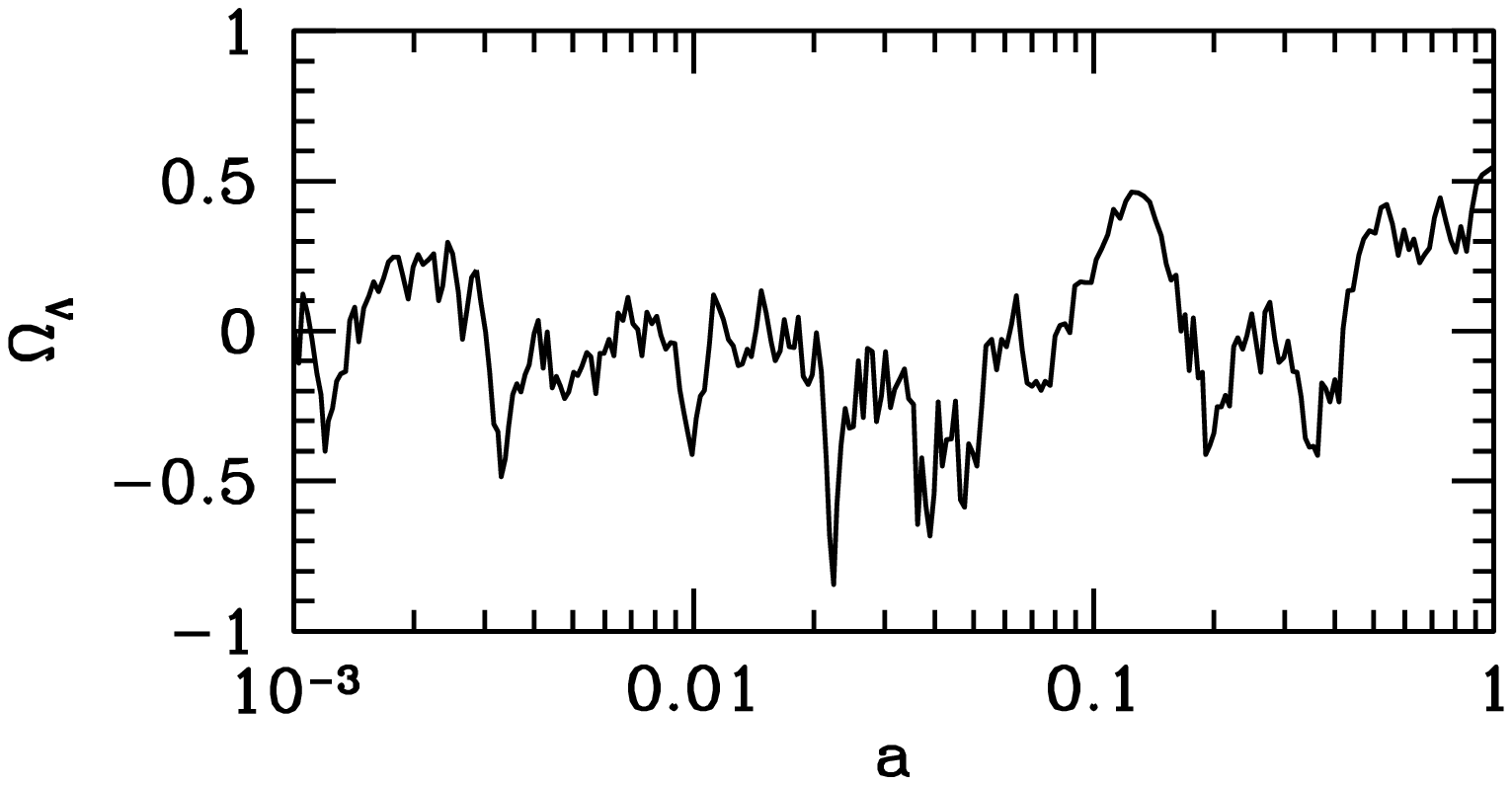}{The ratio of the energy density in cosmological constant to the
total density as a function of scale factor. Here $\alpha=0.02$.}

In half the realizations, $\rho_\Lambda$ will be positive today.
Whether or not it is positive enough to explain the observations 
then becomes a question of probability. 
For $\alpha=0.02$, it clearly is not that
improbable (indeed in the same run, we see another spike in the energy
density at $a\simeq0.1$).

The same qualitative argument applies to the BBN constraint.  In fact
the situation there is even better. Half the time the extra energy
density will be negative, thereby reducing the total energy density in
the universe. This in turn will reduce the predicted abundance of
$^4$He.  There is some disagreement at present as to whether the current
observations agree with the standard cosmological model or
not~\cite{bbn}, with some cosmologists arguing that the observed
abundances are too low. A negative $\rho_\Lambda$ fixes
this problem.

Why have we chosen $\alpha$ to be small? Our choice is in response to a
fundamental incompleteness in our implementation.  
If $\alpha\approx1$, there will
inevitably be times during which the total 
effective
energy density, the sum of
the terms in parenthesis on the right side of \ec{hubble}, goes
negative, thereby invalidating the equation.  
(Whenever this happens, we terminate the run.)
In the next section we offer some thoughts on this problem; here we
simply spell it out.

Figure~\Rf{zvt} shows a history for $\alpha=0.02$ going back to the time
of decoupling.  If we had started earlier, say at the ``Planck time''
$a=10^{-32}$, we would have had only about a 1 in 3 chance of completing
the run without hitting a time at which $\rho_{\rm tot}$ went negative.
Moving $\alpha$ down to $0.01$ evades this problem; for that parameter
choice, very few runs hit a time at which the total energy goes
negative.  However, for $\alpha$ that small, the fluctuations are also
small. Figure~\Rf{alpha} shows a histogram of final values of
$\Omega_\Lambda$ for $6000$ realizations each with $\alpha=0.01$. Only
rarely does the final value of $\Omega_\Lambda$ approach those necessary
to explain the observations.

\Sfig{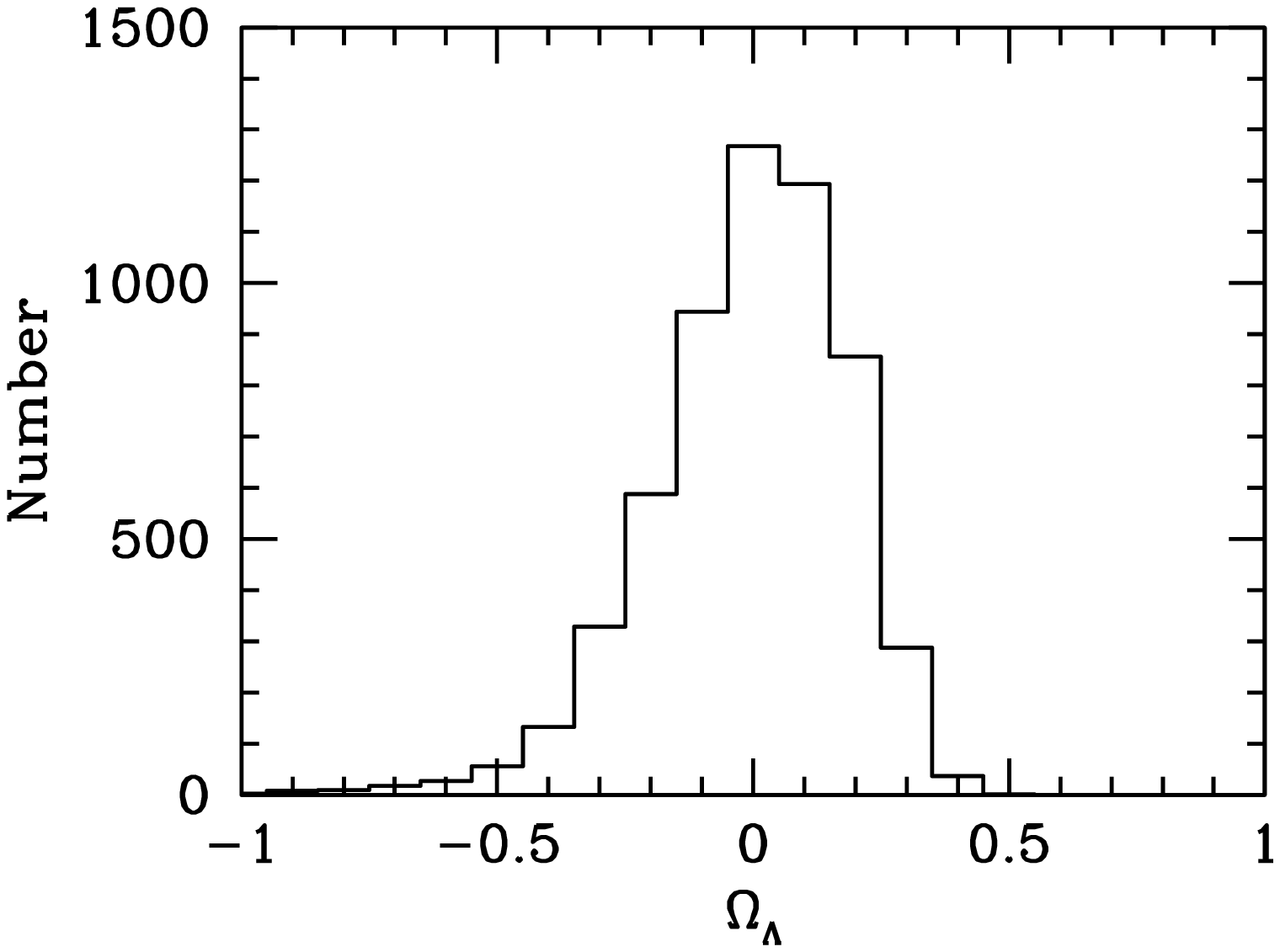}{A histogram of the final value of $\Omega_\Lambda$, the ratio
of $\rho_\Lambda$ to the total density. The dimensionless parameter governing
the fluctuations in $\Lambda$ has been set to $\alpha=0.01$.}

There is therefore a tension: if we push $\alpha$ too low, it becomes
very unlikely that $\rho_\Lambda$ will be large enough today to agree
with observations. If we push $\alpha$ too high, there inevitably comes
a time at which the total energy density in the universe becomes
negative, and the simulation cannot continue.  Of course we are dealing
with probabilities, so for any value of $\alpha$ there is always the
chance that the total energy density remains positive throughout the
history of the universe and the final value of $\rho_\Lambda$ is large
enough to account for observations.  Fortunately, this happens
reasonably often for $\alpha$ in the range $0.01-0.02$.  Nonetheless, we
suspect that we will ultimately have to deal more directly with
possibility that $\rho_{\rm tot}$ goes negative.

\section{Complications}

We can think of two ways to deal with
a negative $\rho_{\rm tot}$
without having to terminate the simulation: 
change the implementation so that this
never occurs or reinterpret $\rho_{\rm tot}$ going to zero (or negative)
so as to give a viable cosmology without having to fine tune $\alpha$.

One approach would be to suppose that $\Lambda$ fluctuates but is positive
semi-definite.  This is the position adopted by Ng and van Dam in
reference~\cite{Ng01}.  There, they argue that the kernel for the
Euclidean gravitational path integral over $\Lambda$ histories takes the form
\be
  e^{ -S_E } \propto
  \exp{\left( \frac{24 \pi^2}{\Lambda} \right)} ,
\ee
in our units.  From this, they argue that the most probable value of
$\Lambda$ is zero and that, if it is not zero, it must be positive.
As they observe, however, this result is peculiar to the assumptions of
Euclidean quantum gravity with all its uncertainties and controversy.
In particular, this result does not seem to follow from causal set theory
or unimodular gravity by themselves and we do not favour it.

Another possibility would be to suppose that the cosmological term 
comes from a decrease in the local energy of one of the matter fields
or gravitational waves.  This is the philosophy of, for example,
Chen and Wu~\cite{ChenWu} who consider a non-fluctuating, but time dependent,
cosmological term $\Lambda(t) \propto a^{-2}(t)$.  
Of course, this supposition is forced upon us if all of Einstein's
equations are to be simultaneously satisfied exactly.
That is,  Einstein's equations---the contracted Bianchi identity in
particular---require that total energy-momentum be conserved.  Thus, in
classical GR, the cosmological term cannot fluctuate without a compensating
fluctuation in the energy-momentum density of one or more of the matter
fields.

As stated in the introduction, our approach has been to solve this problem
by maintaining only the Friedmann equation as exact.
Nevertheless, let us try instead to adapt 
the solution above
to our case.  
Suppose that some matter component, let us take
gravitational waves as a concrete example, is somehow
converted into the energy density of a cosmological term, while the
energy density in every other component (dust, radiation, etc.) is
separately covariantly conserved.  The first law,
$ \frac{d}{dt} \left[\left(\rho_{\rm tot} + p_{\rm tot} \right) a^3 \right]
= a^3 \frac{d p_{\rm tot}}{dt} $,
applied to  these two components becomes:
\begin{equation}
\dot \rho_{\rm grav} = -4H\rho_{\rm grav} - \dot \Lambda(t) \, ,
\end{equation}
where $H$ is the Hubble parameter.  As could be expected, an increase in
$\Lambda$ must lead to a decrease in the energy density in the gravity
waves.  However, for a generic fluctuating $\Lambda(t)$, the cosmological
term might increase enough that the energy density in gravity waves 
becomes negative.
It's not clear how this could be interpreted and it appears that
we would simply exchange one problem with another.  Note that,
in the case of  Chen and Wu~\cite{ChenWu}, this is not a problem.
Their cosmological term decreases {\em monotonically} with the expansion
of the universe.  Thus, we see that this 
solution 
can work with a
fluctuating $\Lambda$ if we demand that $\dot \Lambda \le 0$.
In fact, relaxation processes of this sort have been considered for some
time.  The earliest we are aware of is that of Abbott~\cite{Abbott}.
Recently, there has been
renewed interest in a similar suggestion of Brown and Teitelboim~\cite{BT}
where domains of four-form flux decay spontaneously, relaxing
the effective local value of the cosmological term, see~\cite{GHT} for recent
references.  The difficulty with these proposals is again the ``Why now?''
problem: relaxation rates and/or boundary values must be tuned for any
hope to obtain a viable cosmology.\footnote{Of course, back when Abbott
and Brown and Teitelboim first made their suggestions, there was no compelling
evidence for a non-vanishing cosmological term.  One needed only to make
it small enough.}

All in all, neither of these proposals really seems to address the
central difficulty: 
Within the contexts of causal set theory and unimodular gravity, the
sign of the total
effective
energy density is fundamentally
not constrained.  
We see no good reason to assume either that $\Lambda$
is positive definite or that $\Lambda$ will always decrease.  
Thus, let us
seek instead to understand what happens when $\rho_{\rm tot}$
approaches and perhaps goes through zero.  Our guide will be the
classical theory.  Consider, for example, a dust filled, flat universe
with a negative cosmological constant $\lambda_0 < 0$ (a true constant).
The 0-0 component of Einstein's equations gives us the Friedmann
equation: $3H^2 = \rho_{\rm tot}$.  Meanwhile, the 
$i$-$i$ 
component gives
us the deceleration: $2\frac{\ddot a}{a} = -|\lambda_0| - H^2$.  We see
that, once the matter has red-shifted enough that the total energy
density vanishes, the universe stops expanding and begins to contract.
As it contracts, the energy density in matter once again begins to
exceed the magnitude of the cosmological term and $\rho_{\rm tot}$
never becomes negative.

We expect that something like
this phenomenology will carry over into our case except that,
with a fluctuating cosmological term, it seems likely that the collapse
can reverse itself if the cosmological term later becomes sufficiently
negative a second time.  This is in contrast with
the classical example above where once the universe starts to collapse,
the matter term always dominates and keeps $\ddot a /a$ negative.
In our model, 
however,
we would expect the amplitude of the cosmological term's fluctuations
to
track the matter or radiation energy density in the collapsing
universe.

Of course, none of this follows from the simple evolution ansatz
we have applied  in this paper.  Such detailed dynamical understanding
must await further developments.  Nevertheless, it is reasonable to suppose
that the complete theory will reduce in stages: a full theory
with non-metric structures at the Planck scale and a semi-classical theory
describing metric structures at larger scales.  Furthermore, it is reasonable
that the semi-classical theory will be describable
as some sort of sum-over-histories, where the intermediate states are
three-geometries.  We can envisage the evolution equation we propose as
some sort of classical approximation to propagation in a stochastic potential.
In the sum-over-histories theory, we expect the case 
$\rho_{\rm tot} < 0$
will correspond to a tunneling-type solution.  Our difficulty in handling
$\rho_{\rm tot} < 0$ 
here is, in this sense, no different from the standard problem of finding
an effective, classical description of
barrier penetration.

\section{Conclusion}

It is still too early to understand the full implications
of recent cosmic discoveries that point to dark energy in
the universe. A number of possibilities have previously been explored
in detail, including a non-zero cosmological constant $\Lambda$ and
zero $\Lambda$ with dark energy hidden in a scalar field.

It is also
possible, though, that the measurements are telling us that
we need to modify our understanding of space and time. 
In particular, the
notion that space-time is continuous may be simply an approximation
that breaks down on scales as small as the Planck scale. If so,
drawing on ideas from
causal set theory -- which postulates a discrete space-time --
and unimodular gravity, we have shown that the
cosmological ``constant'' need not be a fixed parameter.  Rather,
it arguably fluctuates
about zero with a magnitude $1/\sqrt{\spv}$, $\spv$ being some measure of
the past four volume.  The amplitude of these fluctuations
is then of the right order of magnitude to explain the dark energy
in the universe. This argument is so general that it would apply
at all times, and, indeed, we expect the energy density in the
cosmological ``constant'' to always be of order the ambient
density in the universe.

In \S IV, we presented a number of issues which inevitably will confront
anyone wishing to implement this idea. Until these issues are resolved,
it will be difficult to make unambiguous, robust
predictions. Nevertheless, one can already see that this theory of a
fluctuating $\Lambda$ differs significantly from most other solutions to
the dark energy problem. Most important for its testability is the
notion that it may have affected the evolution of the universe at early
times. Thus, the primordial generation of perturbations during a
possible inflationary phase;  production of light elements in Big Bang
Nucleosynthesis; acoustic oscillations in the background radiation; and
the evolution of structure at more recent times all may yield clues and
tests of the idea of an everpresent $\Lambda$.

This work was supported by the DOE at 
Fermilab, by NASA grant NAG5-10842 and by NSF Grant PHY-0079251.  It was
also supported at Syracuse University by NSF Grant PHY-0098488 and by an
EPSRC Senior Fellowship at Queen Mary College, University of London.  SD
and RDS would like to acknowledge the hospitality of the Aspen Center
for Physics where their collaboration began, and RDS would like to
acknowledge the hospitality of Goodenough College, London, where part of
this work was done.  PBG would like to acknowledge useful conversations with
J. Moffatt at the University of Toronto and the hospitality of the
Canadian Institute for Theoretical Astrophysics while there.

\newcommand\spr[3]{{\it Physics Reports} {\bf #1}, #2 (#3)}
\newcommand\sapj[3]{ {\it Astrophys. J.} {\bf #1}, #2 (#3) }
\newcommand\sprd[3]{ {\it Phys. Rev. D} {\bf #1}, #2 (#3) }
\newcommand\sprl[3]{ {\it Phys. Rev. Letters} {\bf #1}, #2 (#3) }
\newcommand\np[3]{ {\it Nucl.~Phys. B} {\bf #1}, #2 (#3) }
\newcommand\smnras[3]{{\it Monthly Notices of Royal
	Astronomical Society} {\bf #1}, #2 (#3)}
\newcommand\splb[3]{{\it Physics Letters} {\bf B#1}, #2 (#3)}

\end{document}